# Olfactory receptors for a smell sensor: A comparative study of the electrical responses of rat I7 and human 17-40

**E. Alfinito, J.-F. Millithaler, L. Reggiani**
Dipartimento di Ingegneria dell'Innovazione, Università del Salento,
via Monteroni, Lecce, Italia
CNISM - Consorzio Nazionale Interuniversitario per le Scienze Fisiche della Materia,
via della Vasca Navale, Roma, Italia.

E-mail: eleonora.alfinito@unisalento.it

**Abstract**. In this paper we explore relevant electrical properties of two olfactory receptors (ORs), one from rat OR I7 and the other from human OR 17-40, which are of interest for the realization of smell nanobiosensors. The investigation compares existing experiments, coming from electrochemical impedance spectroscopy, with the theoretical expectations obtained from an impedance network protein analogue, recently developed. The changes in the response due to the sensing action of the proteins are correlated with the conformational change undergone by the single protein. The satisfactory agreement between theory and experiments points to a promising development of a new class of nanobiosensors based on the electrical properties of sensing proteins.

## I. Introduction

The research on smell sensors is widely and long-time explored because of the huge number of possible applications in everyday life (food quality assessment, detection of specific molecules produced in some diseases[1], so as of drugs and toxic material). In order to make these devices more sensitive, easy to use, versatile and low cost, recent advances point to the substitution of their sensitive part, originally constituted by solid-state gas sensors, with organic/biological material[2]. In particular, olfactory receptors (ORs), extracted from different species, have been recently considered among the best candidates to provide the sensitive action in smell sensors of new generation[3].

In this perspective, an European collaboration, now called BOND (bioelectronic olfactory neuron device)[4], has been launched in 2009 and collects different expertises in physics, biology and nanoelectronics. The main goal of the project is to build up an array of nanobiosensors whose active part consists of few kind of ORs, selective on specific odorant molecules, interfaced with nanoelectrodes. The ORs, differently activated by the same odorant compound, should produce a specific odorant response[5], like it happens in vivo. In particular, in vivo, the olfactory receptor activation involves modifications of the receptor topology, which, in turn, produces a cascade of events that culminate in the transmission of the capture information to the brain. In vitro, only the initial part of this chain of events can be reproduced, cutting the process at different stages, with respect to the kind of hybrid system one should develop[6]. Experiments performed with different techniques[7,8] strongly suggest that together with the morphology, the protein electrical properties are modified in this process. Therefore, the OR activation could be monitored by means of electrical measurements.

To date, a set of electrical measurements have been performed[7,9,10] on two olfactory receptors, rat I7 and human 17-40. The measurements evidenced the change of the protein electrical response in the presence of a specific odorant (ligand). In parallel, a theoretical model, able to reproduce the electrical properties of a given protein, has been developed. This model, hereafter called INPA (impedance network protein analogue), describes the protein in terms of an impedance network whose structure reproduces the electrical interaction pathways, consistently with the protein topology. In particular, the model offers the possibility to explore some aspects of the protein morphology and its modifications subsequent to the ligand capture and also to interpret some disputed experimental data within a reliable physical framework.

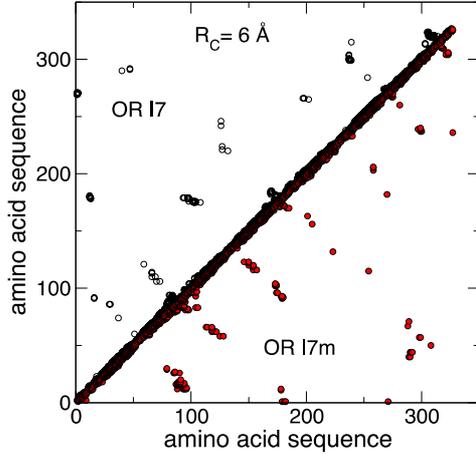

Figure 1: Contact maps for rat OR I7, with $R_C$ = 6Å. The empty circles (left side of diagonal) refer to the native state, OR I7, the full circles (right side of diagonal) refer to the active state, OR I7m.

The aim of this paper is to provide a microscopic interpretation of these experiments and carry out a comparative study of the electrical properties of rat OR I7 and human OR 17-40. To this purpose, the paper is organized as follows. Section II briefly recalls the theoretical model and reports the results together with their physical interpretation. Major conclusions are drawn in Sect. III.

## II. Model and Results

As clearly signaled by experiments, the addition of a specific odorant induces a modification of the electrical response in functionalized samples consisting of smell receptors. As a first attempt, being the proteins investigated without their natural environment, especially the G-protein[11], the main modifications in the electrical response can be attributed to the protein conformational change. Therefore, the model we adopt for describing the experimental outcomes correlates the protein morphology to the protein electrical features. In doing so, the receptor is described by a graph of N nodes where N is the number of the amino acids that constitutes the protein. Then, two nodes are connected if the distance between the corresponding amino acids is not greater than an assigned value $R_C$, the cut-off distance. For $R_C$ varying in the range 3÷70 Å the graph goes from a not connected to a completely connected structure.

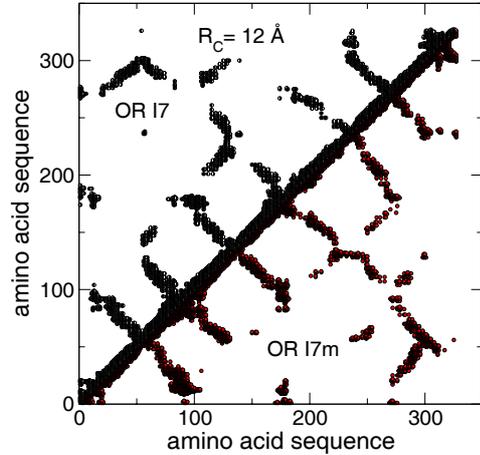

Figure 2: Contact maps for rat OR I7, with $R_C$ =12Å. . The empty circles (left side of diagonal) refer to the native state, OR I7, the full circles (right side of diagonal) refer to the active state, OR I7m.

The distance $R_C$ definitely selects the connectivity of the network, and is here taken as an adjustable parameter of the model. Accordingly, the choice of its value has to be headed by some physical constraints and finally fixed by comparison of numerical calculations with experiments.

As a further step, the graph is turned into an electrical network by substituting each link between two nodes with an RC parallel circuit whose value depends on the distance between the corresponding amino acids. The RC elementary impedance, is given explicitly by[12,13]:

$$Z = \frac{l}{A} \frac{1}{\rho^{-1} + i\varepsilon\varepsilon_0\omega}$$

where $A = \pi(R_C^2 - l^2/4)$, is the cross-sectional area between two spheres of radius $R_C$, each of them centred on one of the two nodes; $l$ is the distance between these centres, ρ is the resistivity, taken to be the same for every amino acid, with the indicative value of ρ $=10^{10}$ Ω m; $i = \sqrt{-1}$ is the imaginary unit, $\varepsilon_0$ is the vacuum permittivity, and ω is the circular frequency of the applied voltage. The relative dielectric constant of the considered couple of nodes (amino acids), ε, is expressed in terms of their intrinsic polarizability[13].

By positioning the input and output electrical contacts, respectively, on the first and last node (corresponding to the first and last amino acid of the protein sequential structure), the network is solved within a linear Kirchhoff scheme and its global impedance spectrum is calculated within the standard frequency range 1 mHz - 100 kHz. For a given interaction radius, the spectrum depends on the network structure.

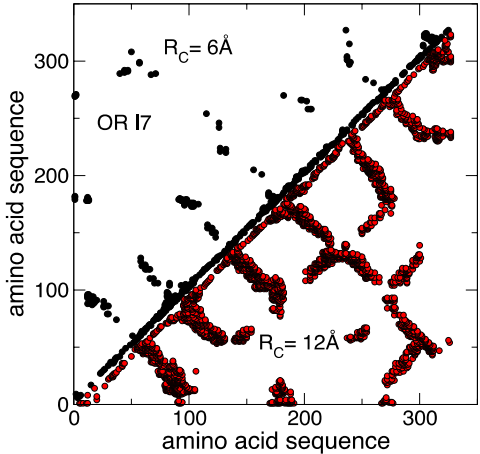

Figure 3: Difference between the number of links in the native and active state of rat OR I7 for $R_C$ = 6 Å (left side of diagonal) and $R_C$ = 12 Å (right side of diagonal).

The change of the protein state, by implying a change in the network structure, leads to a different spectrum of the global impedance. Such a variation of the impedance spectrum is here taken as an estimate of the electrical response, as resulting from the protein conformational change. The investigation of the protein modifications is carried out in the following by means of different tools, such as the analysis of the contact maps, the determination of the protein global resistance, and the shape of the impedance spectra.

## A. Structure and Contact map

The global insights on the protein modification, as induced by the ligand capture, are shown by the contact maps reported in figure 1 and figure 2 for the rat OR I7. In these figures, each point represents a couple of connected amino-acids, i.e. a link, in an assigned configuration (native on the left side of the main diagonal and active on the right side of the main diagonal) and for a given value of the cut-off radius. For both values, $R_C$ = 6 Å (figure 1) and $R_C$ = 12 Å (figure 2), we can observe that the main differences between the native and activate state are located in the middle region, with an increasing of connections for the active state. This is probably due to the protein closing on the binding pocket (there located[14]), subsequent to the ligand capture.

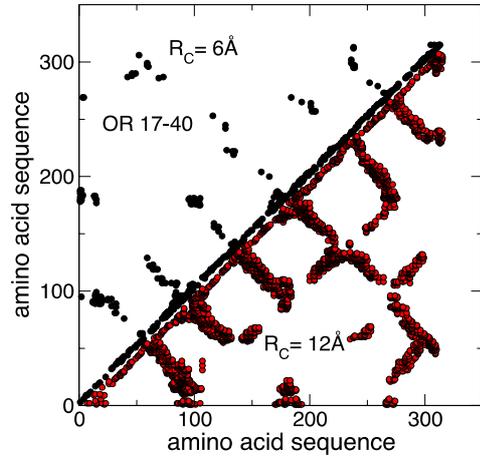

Figure 4: Difference between the number of links in the native and active state of rat OR I7 for $R_C$ = 6 Å (left side of diagonal) and $R_C$ = 12 Å (right side of diagonal).

The difference between the number of links in the native and active state of both the ORs under examination are reported in figure 3 for OR I7 and in figure 4 for OR 17-40. In particular, each figure reports the data at two different $R_C$ values, say 6 Å and 12 Å. By comparing the rat and human protein, we notice that besides the similarities of the two proteins, few but significant differences exist. These differences are also present and amplified in the calculated impedance of the receptors where, more than the absolute difference in the number of links, the link position is of relevance for the determination of the impedance[13].

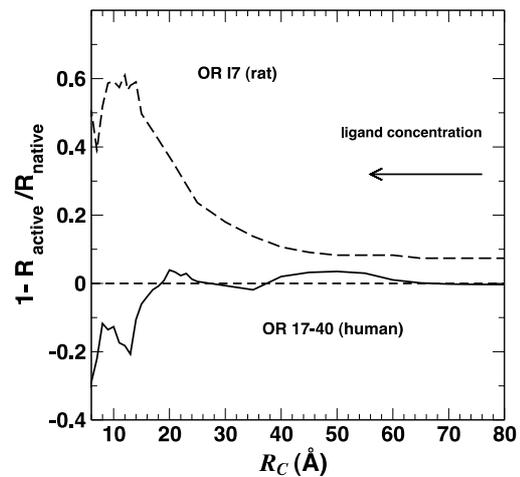

Figure 5: Relative resistance variation for increasing $R_C$ values. Dashed line refers to rat OR I7, continuous line refers to human OR 17-40.

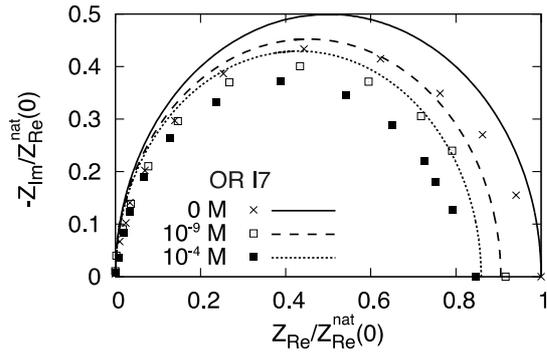 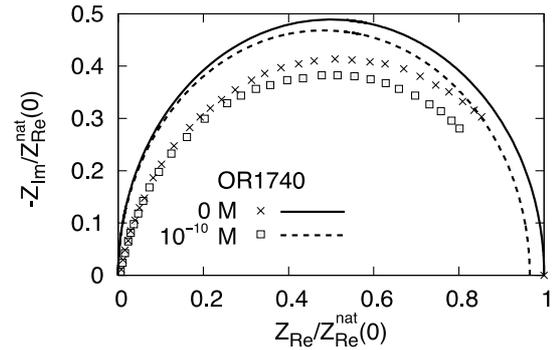

Figure 6: Nyquist plot for rat OR I7 in the absence and in the presence of a specific ligand (octanal). The impedances are normalized to the static value of the native state $Z^{nat}_{Re}(0) = 850\Omega$. Symbols refer to experiments with: crosses refer to no odorant, empty squares to an odorant concentration of $10^{-9}$ M, and full squares to an odorant concentration of $10^{-4}$ M. Lines refer to theoretical results with: continuous curve refers to the native state configuration with $R_C = 50$Å, long dashed line refers to the active state configuration with $R_C = 30$Å, dashed line refers to the active state configuration with $R_C = 25$Å.

Figure 7: Nyquist plot for rat OR 17-40 in the absence and in the presence of the odorant heptanal. The impedances are normalized to the static value of the native state, which, in units of cm$^2$, is $Z^{nat}_{Re}(0) = 33K\Omega$. Symbols refer to experiments with: crosses refer to no odorant, empty squares to an odorant concentration of $10^{-10}$ M. Lines refer to theoretical results with: continuous curve refers to the native state configuration with $R_C = 70$Å, dashed line refers to the active state configuration with $R_C = 46$Å.

## B. Resistance and impedance spectrum

The differences between native and active states are also evidenced in the change of the receptor global resistance, R. Concerning this point, figure 5 shows the shape of the relative variation of R, as function of $R_C$, for both the proteins. The main result reported by this figure is the larger sensitivity of the rat OR I7 with respect to the human OR 17-40: In the former case it is possible to resolve a maximum difference of about 60 % while in the latter case the maximal resolution is of about 20 %. The region of maximal sensitivity is the same for both the proteins, corresponding to $R_C$ in the range $= 6 \div 14$ Å. For the case of OR 17-40, the active state shows values of R which are significantly lower than those of OR I7. Furthermore, for $R_C$ above about 18 Å, calculations evidence for OR 17-40 an inversion of the resistance variation, with the active state becoming less resistive than the native state in two regions of $R_C$ values, $18 \div 24$ Å and $40 \div 60$ Å, respectively. In terms of the electrical network, such an inversion is interpreted as a stronger increase of parallel with respect to series connections. Accordingly, the different behaviour shown by the two receptors signals different peculiarities of the network structure.

The impedance response of single proteins is explored on a wide range of frequencies and the results are given by means of the Nyquist plot. This plot is obtained by drawing the negative imaginary part versus the real part of the global impedance, within a given frequency range (typically from 1 mHz to 100 kHz as in experiments[7,10]). Figure 6 reports the Nyquist plots with the impedance normalized to the static value of its native state for rat OR I7. Symbols refer to EIS measurements[7,15], where the solution resistance is subtracted. Curves refer to theoretical results where the single protein is taken representative of the entire sample. Crosses pertain to experiments carried out in the absence of the specific odorant octanal, empty and full squares pertain to the different concentrations of the odorant as reported in the figure. Within our model, the Nyquist plots are obtained by using as input data the networks corresponding to the native and active states, at the cut-off radius which, according to figure 5, fits the experimental of $Z_{Re}/Z_{nat}(0)$ at the given odorant concentration. In other words, we conjecture the existence of a correlation between the concentration of odorant and $R_C$, due to the modifications undergone by the protein ensemble in the presence of a given concentration of odorant[16]. The agreement between theory and experiments is found to be satisfactory from both a qualitative and quantitative point of view. We remark that the near ideal semicircle shape of the experimental Nyquist plot is well reproduced, thus confirming that the network impedance model behaves closely to a single RC circuit, as expected by the presence of a rather uniform distribution of the time constants associated with the different values of the resistance and capacitance of the links[13].

Figure 7 reports analogous results for the case of human OR 17-40, in the native state and in the active state with a concentration of $10^{-10}$M of the specific odorant heptanal[10]. Again, the agreement between theory and experiments is found to be satisfactory. The scarcity of experimental data and the lack of a certified knowledge of the 3D structures of the considered proteins lead us to consider these results as a first but significant step towards a microscopic modelling of the electrical properties of olfactory receptors.

## III. Conclusions

We have presented a microscopic interpretation of the electrical properties of two OR, the rat I7 and the human 17-40. Both ORs pertain to the huge family of the seven-helices transmembrane receptors, the so called G protein coupled receptors (GPCRs)[17]. Accordingly, they share a similar tertiary structure as well as a similar behaviour in the sensing action. Both proteins are able to bind a specific molecule in the so-called active site, which is placed well inside the protein (among the 3,4,6 helices)[14]. The capture produces a change of the protein conformation (from the native to the active state) then activating a G protein and a biological chain, ending to the brain recognition of the odour. The detection of the conformational change is the first step in the mechanism of detection which should be used by an OR-based nanobiosensor. On the other hand, the detection of the conformational change is not a simple task, especially in vitro, where the cascade of events subsequent the capture, cannot be reproduced. Thus, by looking for the electrical properties of a protein, it is potentially possible to reveal the protein conformational change by measuring the corresponding variations of these properties. In this respect, it is crucial to determine whether this change exists and is sufficiently large to be detectable within the experimental resolution, thus making possible to transduce the sensing action of the protein into an electrical signal. Recent experiments[7,10] have confirmed this possibility by relating the sensing action of the rat OR I7 and the human OR17-40 to the capture of a specific ligand. Measurements were performed on self-assembled multilayers (SAM) substrates on which the proteins were anchored. Then the sensitivity to specific and not-specific odorants and also to the dose-response was tested. The results are found to be promising for the considered proteins, showing a better sensitivity for rat OR I7 than for human OR 17-40[9]. Furthermore, the experiments were accomplished by an electrochemical impedance spectroscopy (EIS) characterization: The conformational change induces a modification of the impedance value, in a wide range of frequencies. The experiments are here microscopically investigated on the basis of an impedance network analogous approach that is found to provide a satisfactory interpretation of available experiments. These results prove the possibility of using proteins as very refined sensors, able to determine, through an electrical signal, the presence and the concentration of the substance to be detected.

This research is carried out within the BOND project under evaluation at the CE within the 7th Program, grant agreement: 228685-2. The active collaboration of Drs. G. Gomila, N. Jaffrezic-Renault, E. Pajot and C. Pennetta is gratefully acknowledged.

Bibliography
[1] A. D'Amico, et al, 2009 Lung Cancer 68 170-176.
[2] E. Alfinito, C. Pennetta and L. Reggiani, 2010 Sensors and Actuators B: Chemical 142 554-558.
[3] W. Göpel, 1998 Sensors and Actuators B: Chemical 52 125-142.
[4] Bioelectronic Olfactory Neuron Device (BOND), Collaborative project FP7-NMP-2008-SMALL-2, GA number 228685.
[5] P. Duchamp-Viret, M.A. Chaput, A. Duchamp, 1999 Science 284 2171-2174.
[6] Q. Liu, et al., 2006 Biosens. Bioelectron. 22, 318-322; J. Minic-Vidic, et al., 2006 Lab. Chip 6 1026-1032.

[7] Y. Hou, et al., 2007 Biosensors & Bioelectronics 22 1550-1555.
[8] Y.-D. Jin, N. Friedman, M. Sheves, T. He and D. Cahen, 2006 PNAS 103 8601-8606; Y.-D. Jin, N. Friedman, M. Sheves, and D. Cahen, 2007 Adv. Funct. Mater. 17 1417- 1428.
[9] J. Vidic, et al., 2008 Lab on a Chip 8 678-688; J. Vidic, et al, 2006 Lab on a Chip 6 1026-1032; G. Levasseur, et al, 2003 Eur. J. Biochem. 270 2905-2912.
[10] I.V. Benilova, et al., 2008 Materials Science & Engineering C 28 633-639.
[11] U. Gether, B. Kobilka, 1998 J. Bio. Chem. 273 17979-17982.
[12] C. Pennetta, et al, 2006 Towards the Realization of Nanobiosensors based on G Protein-Coupled Receptors, in Wiley-VCH Book series on Nanotechnologies for the Life Sciences, vol. 4, Nanodevices for the Life Sciences, ed. by Challa S.S.R. Kumar, Wiley-VCH, Berlin, pp 217-240.
[13] E. Alfinito, C. Pennetta, L. Reggiani, 2009 J. Appl. Phys. 105 084703-1-6.
[14] S. E. Hall, et al., 2004 Chem. Senses 29 595-616; N. Vaidehi, W. B. Floriano, R. Trabanino, S. E. Hall, P. Fred- dolino, E. J. Choi, G. Zamanakos, and W. A. Goddard III, 2002 PNAS 99 12622-12627.
[15] N. Jaffrezic-Renault, private communication.
[16] B. K. Kolbika and X. Deupi, 2007 Trends Pharmacolog. Sci. 28 397-406; O. Miyashita, P.G. Wolynes, and J. N. Onuchic, 2005 J. Phys. Chem. B 109 1959-1969; A. K. Gardino et al. 2009, Cell 139 1109-1118.
[17] R. J. Lefkowitz, 2000 Nature Cell. Bio. 2 E133-E136.